\documentclass[11pt]{article}
\usepackage[blocks,affil-sl]{authblk}
\usepackage[centertags]{amsmath}
\usepackage{amsfonts} \usepackage{amssymb} \usepackage{amsthm}
\allowdisplaybreaks[1]
\usepackage{graphicx}

\def\ii{{\rm i}}

\newcommand{\ft}[2]{{\textstyle\frac{#1}{#2}}}

\begin{document}

\def\thefootnote{\alph{footnote}}
\title{\bf{Resumming instantons in $\mathcal{N}=2^\star$ theories \\
with arbitrary gauge groups \footnote{Proceedings of the XIV Marcel Grossmann Meeting, Rome, Italy, July 12-18, 2015. \newline
E-mail: billo,frau,lerda@to.infn.it; fucito,morales@roma2.infn.it}}}

\author{Marco Bill\`o}
\author{Marialuisa Frau}
\affil{\small{Dipartimento di Fisica, Universit\`a di Torino and I.N.F.N. - Sezione di Torino\newline
Via P. Giuria 1, I-10125 Torino, Italy}}

\author{Francesco Fucito}
\author{Jos\'e F. Morales}
\affil{\small{I.N.F.N. - Sezione di Roma 2 and Dipartimento di Fisica, Universit\`a di Roma Tor Vergata
\newline
Via della Ricerca Scientifica, I-00133 Roma, Italy}}

\author{Alberto Lerda}
\affil{\small{Dipartimento di Scienze e Innovazione Tecnologica, Universit\`a del Piemonte Orientale
\newline
and I.N.F.N. - Gruppo Collegato di Alessandria - Sezione di Torino
\newline
Viale T. Michel  11, I-15121 Alessandria, Italy}}

\date{\vspace{-5ex}}
\maketitle

\begin{abstract}
\noindent
We discuss the modular anomaly equation satisfied by the the prepotential of 4-dimensional 
$\mathcal{N}=2^\star$ theories and show that its validity 
is related to $S$-duality.
The recursion relations that follow from the modular anomaly equation 
allow one to write the prepotential in terms of (quasi)-modular forms, 
thus resumming the instanton contributions. 
These results can be checked against the microscopic multi-instanton calculus in the case of 
classical algebras, but are valid also for the exceptional $E_{6,7,8}$, $F_4$ and $G_2$
algebras, where direct computations are not available.

\end{abstract}

\setcounter{footnote}{0}
\def\thefootnote{\arabic{footnote}}

\vspace{0.4cm}
\section{Introduction}
These proceedings are based on the papers \cite{Billo':2015ria} where we
studied $\mathcal{N}=2^\star$ SYM theories with a gauge algebra
$\mathfrak{g}\in\{{\tilde A}_r,B_r,C_r,D_r,E_{6,7,8},F_4,G_2\}$, extending previous results
obtained in \cite{Billo:2013fi} for the unitary groups.%
\footnote{Here and in the following we denote by ${\tilde A}_r$ the algebra of the unitary group $\mathrm{U}(r+1)$.} 
Our motivation is to shed light on the general structure of $\mathcal{N}=2^\star$ SYM theories 
at low energy and 
show that the constraints imposed by $S$-duality take the form of a recursion relation which 
allows one to determine the prepotential at a non-perturbative level and resum
all instanton contributions.

The $\mathcal{N}=2^\star$ theories arise as deformations of the $\mathcal{N}=4$ theories when
the adjoint hypermultiplet acquires a mass $m$. Their low-energy effective dynamics is entirely 
encoded in the prepotential, which we denote as $F^\mathfrak{g}$ and which is a holomorphic 
function of the coupling constant
\begin{equation}
\label{deftau}
 \tau=\frac{\theta}{2\pi}+\mathrm{i}\frac{4\pi}{g^2}~,
\end{equation}
and of the vacuum expectation value $a$ of the scalar field in the adjoint vector multiplet. For definiteness, we
take $a$ along the Cartan directions of $\mathfrak{g}$, namely
\begin{equation}
a =\mathrm{diag}\big(a_1,a_2,\cdots,a_\mathfrak{r}\big) 
\end{equation}
where 
$\mathfrak{r}=\mathrm{rank}(\mathfrak{g})$.\footnote{The special unitary case, corresponding to the algebra $A_r$ is recovered by simply imposing the tracelessness condition on $a$.}
To treat all algebras simultaneously it is convenient to introduce the parameter
\begin{equation}
 n_{\mathfrak{g}}=\frac{\alpha_{\mathrm{L}}\cdot \alpha_{\mathrm{L}}}{\alpha_{\mathrm{S}}\cdot \alpha_{\mathrm{S}}}
\label{ng}
\end{equation}
where $\alpha_{\mathrm{L}}$ and $\alpha_{\mathrm{S}}$ are, respectively, the long and the short 
roots of $\mathfrak{g}$. For the root system $\Psi_{\mathfrak{g}}$, we follow the standard conventions 
\cite{Billo':2015ria} (see also the Appendix), so that
\begin{equation}
 \begin{aligned}
 n_{\mathfrak{g}}&=1\quad\mathrm{for}~\mathfrak{g}={\tilde A}_r,D_r,E_{6,7,8}~,\\
 n_{\mathfrak{g}}&=2\quad\mathrm{for}~\mathfrak{g}=B_r,C_r,F_4~,\\
 n_{\mathfrak{g}}&=3\quad\mathrm{for}~\mathfrak{g}=G_2~.
 \end{aligned}
\label{ngvalues}
\end{equation}
Using this, one finds that
\begin{equation}
 F^\mathfrak{g}(\tau,a)= n_{\mathfrak{g}}\mathrm{i}\pi\tau a^2+ f^\mathfrak{g}(\tau,a)
\end{equation}
where the first term is the classical contribution while $f^\mathfrak{g}$ is the quantum part. The
latter has a $\tau$-independent one-loop term
\begin{equation}
\label{F1loop}
 f^\mathfrak{g}_\mathrm{1-loop} = \frac{1}{4}\sum_{\alpha\in\Psi_{\mathfrak{g}}} \Bigl[
 -(\alpha\cdot a)^2 \log\left(\frac{\alpha\cdot a}{\Lambda}\right)^2 +
 (\alpha\cdot a+ m)^2 \log\Bigl(\frac{\alpha\cdot a+m}{\Lambda}\Bigr)^2\,\Bigr]
\end{equation}
where $\Lambda$ is an arbitrary scale, and a series of non-perturbative corrections at instanton 
number $k$ proportional to $q^k$, where $q = \exp(2\pi\ii \tau)$. 

The quantum prepotential can be expanded in even powers of $m$ as
\begin{equation}
\label{fexpm}
f^\mathfrak{g}(\tau,a)=\sum_{n\geq 1}^{\phantom{.}}f_n^\mathfrak{g}(\tau,a)
\end{equation}
with $f_n^\mathfrak{g}$ proportional to $m^{2n}$. The first coefficient $f_1^\mathfrak{g}$
receives only a contribution at one-loop and, thus, 
is independent of $\tau$. 
For $n>1$, instead, the coefficients $f_n^\mathfrak{g}$ receive contributions also from the instanton sectors.
When $\mathfrak{g}\in\{{\tilde A}_r,B_r,C_r,D_r\}$, these 
non-perturbative terms can be computed using localization techniques \cite{Nekrasov:2002qd,Bruzzo:2002xf,Shadchin:2004yx,Billo:2012st} as we will show in Section~\ref{secn:localization}, but for the exceptional algebras they have to be derived with 
other methods. As a by-product, our analysis provides also an explicit derivation of all instanton 
contributions to the prepotential for the exceptional algebras $E_{6,7,8}$, $F_4$ and $G_2$, 
at least for the first few values of $n$. The key ingredient for this is $S$-duality.

\section{S-duality}
In $\mathcal{N}=4$ SYM theories with gauge algebra $\mathfrak{g}$, the duality group is generated by
\begin{equation}
S=\begin{pmatrix}
0&-1/\sqrt{n_{\mathfrak{g}}} \\
\sqrt{n_{\mathfrak{g}}} &0
\end{pmatrix}\quad\mbox{and}\quad
T=\begin{pmatrix}
\,1\,&\,1\, \\
\,0\, &\,1\,
\end{pmatrix}~,
\label{ST}
\end{equation}
which, on the coupling constant $\tau$, act projectively as follows
\begin{equation}
\label{Stau}
S(\tau) = - \frac{1}{n_\mathfrak{g}\tau}\quad\mbox{and}\quad
T(\tau) = \tau+1~.
\end{equation}
The matrices (\ref{ST}) satisfy the constraints
\begin{equation}
S^2=-1\quad\mbox{and}\quad
(ST)^{p_{\mathfrak{g}}}=-1\quad\mbox{with}~~n_{\mathfrak{g}}
=4\cos^2\big(\ft{\pi}{p_{\mathfrak{g}}}\big)~,
\end{equation}
and generate a subgroup of SL(2,$\mathbb{R}$) which is known as the Hecke group H($p_{\mathfrak{g}}$).
For the simply laced algebras, {\it{i.e.}} $n_{\mathfrak{g}}=1$, we have $p_{\mathfrak{g}}=3$, and the duality group H(3) is just the modular group $\Gamma=\mathrm{SL}(2,\mathbb{Z})$. For the non-simply laced algebras, the duality groups H(4) and H(6), corresponding respectively 
to $n_{\mathfrak{g}}=2$ and $n_{\mathfrak{g}}=3$, are clearly different from the modular group 
but contain subgroups which are also congruence subgroups of $\Gamma$. 
Indeed, one can show that the following H($p_{\mathfrak{g}}$) elements
\begin{equation}
V=STS=\begin{pmatrix}
-1&\,\,0\, \\
n_{\mathfrak{g}} &-1
\end{pmatrix}\quad\mbox{and}\quad
T=\begin{pmatrix}
\,1\,&\,1\, \\
\,0\, &\,1\,
\end{pmatrix}
\label{STn}
\end{equation}
generate
\begin{equation}
\Gamma_0(n_{\mathfrak{g}}) = \Bigg\{
\begin{pmatrix}
~a~&~b~ \\
~c~&~d~
\end{pmatrix} \in \Gamma~: ~c=0~~\mbox{mod}~n_{\mathfrak{g}}
\Bigg\}\subset \Gamma~.
\label{gamma0n}
\end{equation}
As we will see, the modular forms of $\Gamma_0(n_{\mathfrak{g}})$, which are known and classified, and have a simple behavior also under S-duality, play an important role for the $\mathcal{N}=2^\star$ SYM theories.\footnote{It is interesting to observe
that for $n_{\mathfrak{g}}=3$, the matrices $T$ and $V^2$ generate the subgroup $\Gamma_1(3)$, 
whose modular forms play a role in the $\mathcal{N}=2$ SYM theory with gauge group SU(3) and six fundamental hypermultiplets \cite{Ashok:2015cba}.}

Another important feature is that the duality transformations exchange electric states of the theory 
with gauge algebra $\mathfrak{g}$ with magnetic states of the theory with the GNO dual 
algebra $\mathfrak{g}^{\!\vee}$, which is obtained from $\mathfrak{g}$ by exchanging
(and suitably rescaling) the long and the short roots \cite{Goddard:1976qe}.
The correspondence between $\mathfrak{g}$ and $\mathfrak{g}^{\!\vee}$ is given in the following table
\begin{equation*}
\begin{tabular}{c|ccccccc}
~~$\mathfrak{g}\phantom{\Big|}$~~&~~${\tilde A}_r$~~&~~$B_r$~~&~~$C_r$~~&~~$D_r$~~&~~$E_{6,7,8}$~~&~~$F_4$~~
&~~$G_2$~~\\
\hline
~~$\mathfrak{g}^{\!\vee}\phantom{\Big|}$~~&~~${\tilde A}_r$~~&~~$C_r$~~&~~$B_r$~~&~~$D_r$~~&~~$E_{6,7,8}$~~&~~$F^\prime_4$~~&~~$G^\prime_2$~~\\
\end{tabular}
\end{equation*}
where for $F_4$ and $G_2$, the ${}^\prime$ in the last two columuns means that the dual root 
systems are equivalent to the original ones up to a rotation.

This duality structure remains and gets actually enriched
when the $\mathcal{N}=4$ SYM theories are deformed into 
the corresponding $\mathcal{N}=2^\star$ ones. Here the $S$ transformation (\ref{ST}) relates the electric variable $a$ of the $\mathfrak{g}$ theory
with the magnetic variable $a_{\mathrm{D}} $ of the dual $\mathfrak{g}^{\!\vee}$ theory
\begin{equation}
\label{defaD}
a_{\mathrm{D}} \,\equiv\,\frac{1}{2 \pi\ii n_{\mathfrak{g}}}
\frac{\partial F^{\mathfrak{g}^{\!\vee}}}{\partial a}= 
 \tau \Big(a+\frac{1}{2\pi\ii n_{\mathfrak{g}}\tau}\frac{\partial f^{\mathfrak{g}^{\!\vee}}}{\partial a}\Big)~,
\end{equation} 
according to
\begin{equation}
S \begin{pmatrix}
a_{\mathrm{D}} \\
a
\end{pmatrix}\,
=
\begin{pmatrix}
0&-1/\sqrt{n_{\mathfrak{g}}} \\
\sqrt{n_{\mathfrak{g}}}&0
\end{pmatrix}\,\begin{pmatrix}\
a_{\mathrm{D}} \\
a
\end{pmatrix}
=\begin{pmatrix}
-a/\sqrt{n_{\mathfrak{g}}} \\
\sqrt{n_{\mathfrak{g}}}\, a_{\mathrm{D}}
\end{pmatrix}~.
\label{saad}
\end{equation}
In other words, the $S$ transformation exchanges the description based on $a$ with its Legendre-transformed one, based on $a_{\mathrm{D}}$:
\begin{equation}
S[F^\mathfrak{g}] = \mathcal{L}[F^{\mathfrak{g}^{\!\vee}}]~,
\label{SisL}
\end{equation} 
where the Legendre transform is defined as 
\begin{equation}
\label{Legendre}
\begin{aligned}
\mathcal{L}\big[F^{\mathfrak{g}^{\!\vee}}\big]
&\equiv F^{\mathfrak{g}^{\!\vee}} - a\cdot\frac{\partial F^{\mathfrak{g}^{\!\vee}}}{\partial a}
=-n_{\mathfrak{g}}\pi\ii\tau a^2 - a\cdot\frac{\partial f^{\mathfrak{g}^{\!\vee}}}{\partial a}+f^{\mathfrak{g}^{\!\vee}} ~.
\end{aligned}
\end{equation}
Thus, as is clear from (\ref{SisL}), $S$-duality is not a symmetry of the effective theory since it changes the gauge algebra; nevertheless, as we shall see, it is powerful enough to constrain the form of the prepotential at the non-perturbative level.

\section{The modular anomaly equation}
If one uses eq.s (\ref{Stau}), (\ref{defaD}) and (\ref{saad}) to evaluate $S[F^\mathfrak{g}]$, 
the requirement (\ref{SisL}) can be recast in the following form:
\begin{equation}
\label{SisL1}
f^\mathfrak{g}\Bigl(\!-\ft{1}{n_\mathfrak{g}\tau},\sqrt{n_{\mathfrak{g}}}a_D\Bigr) = 
\frac{1}{4\pi\ii n_{\mathfrak{g}}\tau} \Big(\frac{\partial f^{\mathfrak{g}^{\!\vee}}}{\partial a}\Big)^2
+ f^{\mathfrak{g}^{\!\vee}}
\end{equation} 
where the r.h.s. is evaluated in $\tau$ and $a$. 

Eq. (\ref{SisL1}) can be solved assuming that the coefficients $f_n$ in the 
mass expansion (\ref{fexpm}) of the quantum prepotential
depend on $\tau$ only through quasi-modular forms of $\Gamma_0(n_{\mathfrak{g}})$.
The ring of these quasi-modular forms is generated by
\begin{equation}
\label{modformsgen}
\begin{aligned}
\left\{E_2, E_4, E_6\right\}~~~& 
\mbox{for}~n_{\mathfrak{g}} = 1~,\\
\left\{E_2, H_2, E_4, E_6\right\}~~~ 
& \mbox{for}~n_{\mathfrak{g}} = 2,3~,
\end{aligned}
\end{equation}
where $E_n(\tau)$ are the Eisenstein series while 
\begin{equation}
H_2(\tau)=\Bigg[\Big(\frac{\eta^{n_\mathfrak{g}}(\tau)}{\eta(n_\mathfrak{g}\tau)}\Big)^{\lambda_\mathfrak{g}}+\lambda_\mathfrak{g}^{n_\mathfrak{g}}
\Big(\frac{\eta^{n_\mathfrak{g}}(n_\mathfrak{g}\tau)}{\eta(\tau)}\Big)^{\lambda_\mathfrak{g}}
\Bigg]^{1-\frac{1}{n_{\mathfrak{g}}}}
\end{equation}
where $\eta$ is the Dedekind $\eta$-function and $\lambda_{\mathfrak{g}}= n_{\mathfrak{g}}^{\frac{6}{n_{\mathfrak{g}}(n_{\mathfrak{g}}-1)}}$. Thus, $\lambda_{\mathfrak{g}}=8,3$ for
$n_{\mathfrak{g}}=2,3$ respectively.
All these forms admit a Fourier expansion in terms of the instanton weight $q$, which
starts as $1 + O(q)$. This means that their perturbative part is just 1.
Being able to express the prepotential in terms of quasi-modular forms entails resumming 
its istanton expansion.

The modular forms (\ref{modformsgen}) transform in a simple way also under $S$; in fact
\begin{subequations}
\begin{align}
&H_2\big(\!-\ft{1}{n_\mathfrak{g}\tau}\big)=-\big(\sqrt{n_{\mathfrak{g}}}\,\tau\big)^2 H_2~,\\
&E_2\big(\!-\ft{1}{n_\mathfrak{g}\tau}\big)=\big(\sqrt{n_{\mathfrak{g}}}\,\tau\big)^2
\Bigl[E_2+(n_{\mathfrak{g}}-1)
H_2+\delta\Bigr]~,\\
&E_4\big(\!-\ft{1}{n_\mathfrak{g}\tau}\big)=\big(\sqrt{n_{\mathfrak{g}}}\,\tau\big)^4\Bigl[E_4
+5(n_{\mathfrak{g}}-1)H_2^2+(n_{\mathfrak{g}}-1)(n_{\mathfrak{g}}-4)E_4
\Bigr]~,\\
&E_6\big(\!-\ft{1}{n_\mathfrak{g}\tau}\big)=\big(\sqrt{n_{\mathfrak{g}}}\,\tau\big)^6
\Bigl[E_6+\ft{7}{2}(n_{\mathfrak{g}}-1)(3n_{\mathfrak{g}}-4)H_2^3
\notag\\
&\qquad\qquad\qquad\qquad\qquad~~~
-\ft{1}{2}(n_{\mathfrak{g}}-1)(n_{\mathfrak{g}}-2)(7E_4\,H_2+2E_6)\Bigr]~,
\end{align}
\label{stransf}
\end{subequations}
where $\delta=\frac{6}{\pi\ii\tau}$.
Thus a quasi-modular form of $\Gamma_0(n_{\mathfrak{g}})$ with weight $w$ is mapped under $S$ to 
a form of the same weight with a prefactor $(\sqrt{n_{\mathfrak{g}}}\tau)^w$, up to the 
$\delta$-shift introduced by $E_2$. 

Suppose moreover that the coefficients $f_n^{\mathfrak{g}}$ enjoy the following property:
\begin{equation}
f_n^\mathfrak{g}\Bigl(\!-\ft{1}{n_\mathfrak{g}\tau},a\Bigr)
= \big(\sqrt{n_{\mathfrak{g}}}\,\tau\big)^{2n-2} \,
\left.f_n^{\mathfrak{g}^{\!\vee}}(\tau,a)\right|_{E_2\to E_2 + \delta}~.
\label{fng}
\end{equation}
If we use this relation in the l.h.s. of eq. (\ref{SisL1}) and take into account eq. (\ref{defaD}), upon
formally expanding in $\delta$ we obtain
 \begin{equation}
\frac{\partial f^{\mathfrak{g}^{\!\vee}}}{\partial E_2}+\frac{1}{24\,n_{\mathfrak{g}}}
\frac{\partial f^{\mathfrak{g}^{\!\vee}}}{\partial a}\cdot
\frac{\partial f^{\mathfrak{g}^{\!\vee}}}{\partial a}=0~;
\label{diffeq}
\end{equation}
of course, since we considered a generic case, we could have equivalently written 
it in terms of $f^\mathfrak{g}$.
This equation governs the appearance in the quantum prepotential of terms containing the second
Eisenstein series $E_2$, which is the only source of a \emph{quasi}-modular behaviour.
Using the mass expansion (\ref{fexpm}), this ``modular anomaly'' equation becomes a recursion relation
\begin{equation}
\label{recrel}
\frac{\partial f_n^\mathfrak{g}}{\partial E_2}=-\frac{1}{24\,n_{\mathfrak{g}}}
\sum_{\ell=1}^{n-1} 
\frac{\partial f_\ell^\mathfrak{g}}{\partial a}\cdot
\frac{\partial f_{n-\ell}^\mathfrak{g}}{\partial a}~.
\end{equation}

\subsection{Exploiting the modular anomaly}
\label{subsec:explmodan}
Starting from $f_1^\mathfrak{g}$, we can use the relation (\ref{recrel}) to determine the parts of the 
$f_n^\mathfrak{g}$'s which explicitly contain $E_2$. The remaining terms of $f_n^\mathfrak{g}$ are strictly modular; we fix them by comparison with the result of the explicit computation of $f_n^\mathfrak{g}$ 
via localization techniques, when available, up to instanton order $(d_{2n-2}-1)$
where $d_{2n-2}$ is the number of independent modular forms of weight $(2n-2)$. 
Once this is done, the resulting expression 
is valid at \emph{all} istanton orders. We stress that the modular anomaly implements a symmetry 
requirement and does not eliminate the need of a dynamical input; yet it is extremely powerful 
as it greatly reduces it.       

The mass expansion of the one-loop prepotential (\ref{F1loop}) reads
\begin{equation}
\label{F1loopex}
 f^\mathfrak{g}_{\mathrm{1-loop}} = 
 \frac{m^2}{4} \sum_{\alpha\in\Psi_{\mathfrak{g}}}  \log\left(\frac{\alpha\cdot a}{\Lambda}\right)^2
- \sum_{n=2}^\infty  \frac{m^{2n}}{4n(n-1)(2n-1)} 
\Bigl(L^\mathfrak{g}_{2n-2}+S^\mathfrak{g}_{2n-2}\Bigr)
\end{equation}
where we introduced the sums
\begin{equation}
\begin{aligned}
L^\mathfrak{g}_{n;\,m_1\cdots\, m_\ell}&=
\sum_{\alpha\in\Psi^{\mathrm{L}}_{\mathfrak{g}}}~\sum_{\beta_1\not=\cdots\beta_\ell\in
\Psi_\mathfrak{g}(\alpha)}
\frac{1}{(\alpha\cdot a)^n(\beta_1\cdot a)^{m_1}\cdots(\beta_\ell\cdot a)^{m_\ell}}~,\\
S^\mathfrak{g}_{n;\,m_1\cdots\, m_\ell}&=
\sum_{\alpha\in\Psi^{\mathrm{S}}_{\mathfrak{g}}}~
\sum_{\beta_1\not=\cdots\beta_\ell\in\Psi^\vee_\mathfrak{g}(\alpha)}
\frac{1}{(\alpha\cdot a)^n(\beta^\vee_1\cdot a)^{m_1}
\cdots(\beta^\vee_\ell\cdot a)^{m_\ell}}~,
\end{aligned}
\label{sumsLS}
\end{equation}
which are crucial in expressing the results of the recursion procedure.
Here $\Psi^{\mathrm{L}}_{\mathfrak{g}}$ and $\Psi^{\mathrm{S}}_{\mathfrak{g}}$ denote, respectively, 
the sets of long and short roots of $\mathfrak{g}$, and for any root $\alpha$ we have defined
\begin{equation}
\begin{aligned}
\label{defPsialpha}
\Psi_\mathfrak{g}(\alpha)
&=\left\{\beta\in\Psi_{\mathfrak{g}} \, : \,\alpha^{\!\vee}\cdot\beta=1\right\}~,
\\
\Psi_\mathfrak{g}^{\vee}(\alpha)
&=\left\{\beta\in\Psi_{\mathfrak{g}} \, :\,\alpha\cdot\beta^{\vee}=1\right\}
\end{aligned}
\end{equation}
with $\alpha^{\!\vee}$ being the coroot of $\alpha$.
For the ADE algebras ($n_\mathfrak{g} = 1$) all roots are long and only the 
sums of type $L^\mathfrak{g}_{n;\,m_1\cdots\, m_\ell}$ exist. Thus,
in all subsequent formul\ae\, the sums $S^\mathfrak{g}_{n;\,m_1\cdots\, m_\ell}$ are to be set 
to zero in these cases.  

The initial condition for the recursion relation (\ref{recrel}) is $f_1^\mathfrak{g}$. Since this 
receives contribution only at one-loop, it can be read from the term of order $m^2$ in eq. (\ref{F1loopex}). Then, the first step of the recursion reads
\begin{equation}
\label{recf2}
\frac{\partial f^\mathfrak{g}_2}{\partial E_2} = -\frac{1}{24 n_{\mathfrak{g}}} 
\frac{\partial f^\mathfrak{g}_1 }{\partial a}\cdot \frac{\partial f^\mathfrak{g}_1 }{\partial a}
= -\frac{m^4}{96 n_{\mathfrak{g}}}
\sum_{\alpha,\beta\in\Psi_{\mathfrak{g}}} \frac{\alpha\cdot\beta}{(\alpha\cdot a)(\beta\cdot a) }
= -\frac{m^4}{24} \left(L^\mathfrak{g}_2 + \frac{1}{n_{\mathfrak{g}}} S^\mathfrak{g}_2\right)
\end{equation} 
where the last equality follows from the properties of the root system $\Psi_{\mathfrak{g}}$.

For $n_{\mathfrak{g}}=1$ there are no forms of weight 2 other than $E_2$ (see (\ref{modformsgen})), 
and thus $f^\mathfrak{g}_2$ only depends on $E_2$. 
For $n_{\mathfrak{g}}=2,3$, instead, $f^\mathfrak{g}_2$ may contain also the 
other modular form of degree $2$ that exists in these cases, namely $H_2$ .
The coefficient of $H_2$ in $f_2^{\mathfrak{g}}$ 
is fixed by matching the perturbative term with the $m^4$ term in eq. (\ref{F1loopex}), namely 
$-\ft{m^4}{24}(L^\mathfrak{g}_2 + S^\mathfrak{g}_2)$. 
In this way we completely determine the expression of 
$f_2^{\mathfrak{g}}$. The process can be continued straightforwardly to higher orders in the mass expansion, though of course the structure gets rapidly more involved. In \cite{Billo':2015ria}  we gave the results up to order $m^{10}$ for the simply-laced algebras, and up to $m^8$ for the non simply-laced ones. 
Here, for the sake of brevity we only report the results up to order $m^6$, namely $f_2^\mathfrak{g}$ and
$f_3^\mathfrak{g}$ :
\begin{eqnarray}
f_2^\mathfrak{g}&=&
-\frac{m^4}{24}E_2\,L_2^\mathfrak{g}-\frac{m^4}{24n_{\mathfrak{g}}}\Bigl[E_2+(n_{\mathfrak{g}}-1)H_2\Bigr]\,S_2^\mathfrak{g}~,\label{res2} \\
f_3^\mathfrak{g}&=&-\frac{m^6}{720}\Bigl[5E_2^2+E_4\Bigr]\,L_4^\mathfrak{g}
-\frac{m^4}{576}\Bigl[E_2^2-E_4\Bigr]\,L_{2;11}^\mathfrak{g}\nonumber\\
&&-\frac{m^6}{720n_{\mathfrak{g}}^2}\Bigl[5E_2^2+E_4+10(n_{\mathfrak{g}}-1)E_2 H_2\nonumber\\
&&\qquad\qquad+5n_{\mathfrak{g}}(n_{\mathfrak{g}}-1)H_2^2
+(n_{\mathfrak{g}}-1)(n_{\mathfrak{g}}-4)E_4\Bigr]\,S_4^\mathfrak{g}\label{res3}\\
&&-\frac{m^6}{576n_{\mathfrak{g}}^2}\Bigl[E_2^2-E_4+2(n_{\mathfrak{g}}-1)E_2 H_2\nonumber\\
&&\qquad\qquad+(n_{\mathfrak{g}}-1)(n_{\mathfrak{g}}-6)H_2^2-(n_{\mathfrak{g}}-1)(n_{\mathfrak{g}}-4)E_4\Bigr]
\,S_{2;11}^\mathfrak{g}~.\nonumber
\end{eqnarray}
Consistency requires that the $f_n^\mathfrak{g}$'s obtained from the recursion procedure satisfy 
eq.~(\ref{fng}). For the ADE algebras ($n_{\mathfrak{g}}=1$), using the modular
properties of the Eisenstein series, it is not difficult to show that they do. On the other hand, 
for the non-simply laced algebras ($n_{\mathfrak{g}}=2,3$), 
using the properties of the root systems, one can prove that 
\begin{equation}
\begin{aligned}
L_{n;m_1\cdots m_\ell}^\mathfrak{g}
&
= \Big(\frac{1}{\sqrt{n_{\mathfrak{g}}}}\Big)^{n+m_1+\cdots+m_\ell}
\,S_{n;m_1\cdots m_\ell}^{\mathfrak{g}^{\!\vee}}~,
\\
S_{n;m_1\cdots m_\ell}^\mathfrak{g}
&
= \big(\sqrt{n_\mathfrak{g}}\big)^{n+m_1+\cdots+m_\ell}
\,L_{n;m_1\cdots m_\ell}^{\mathfrak{g}^{\!\vee}}~.
\end{aligned}
\label{LSa}
\end{equation}
These duality relations, together with the modular transformations (\ref{stransf}), 
ensure that the expressions in eq.s~(\ref{res2}) and (\ref{res3}), as well as those arising
at higher mass orders, indeed obey eq.~(\ref{fng}).

\subsection{One-instanton contributions}
By considering the instanton expansion of the modular forms appearing in the expression 
of the $f_n^\mathfrak{g}$'s, one can see that at the one-instanton order, {\it{i.e.}} at order $q$, 
the only remaining terms involve the sums of type $L^\mathfrak{g}_{2;1\cdots 1}$. In fact it can be argued from the recursion relation that this is the case at any 
order in the mass expansion. Thus, the one-instanton prepotential reads
\begin{align}
F^\mathfrak{g}_{k=1}&=m^4\sum_{\ell\geq 0} \frac{m^{2\ell}}{\ell!}\,
L^\mathfrak{g}_{2;{\underbrace{\mbox{\scriptsize{1\ldots1}}}_{\mbox{\scriptsize{$\ell$}}}}}\notag
\\
&=\sum_{\alpha\in\Psi^{\mathrm{L}}_\mathfrak{g}}\frac{m^4}{(\alpha\cdot a)^2}\,\sum_{\ell\geq 0} \frac{m^{2\ell}}{\ell!}
\!\!
\sum_{\beta_1\not=\cdots\not=\beta_\ell\in\Psi_\mathfrak{g}(\alpha)} \frac{1}{(\beta_1\cdot a)
\cdots(\beta_\ell\cdot a)}\label{F1}\\
&=\sum_{\alpha\in\Psi^{\mathrm{L}}_\mathfrak{g}}\frac{m^4}{(\alpha\cdot a)^2}\prod_{\beta\in\Psi_\mathfrak{g}(\alpha)}\left(1+
\frac{m}{\beta\cdot a}\right)\notag
\end{align}
where the intermediate step follows from the definition (\ref{sumsLS}) of the  
sums $L^\mathfrak{g}_{2;1\cdots1}$. The number of factors in the product above
is given by the order of $\Psi_\mathfrak{g}(\alpha)$. When $\alpha$ is
a long root, this is $(2h_\mathfrak{g}^{\!\vee}-4)$ where $h_\mathfrak{g}^{\!\vee}$ is the dual 
Coxeter number of $\mathfrak{g}$ (see the Appendix).
Thus, in (\ref{F1}) the highest power of the mass is $m^{2h_\mathfrak{g}^{\!\vee}}$. 
This is precisely the only term which survives in the decoupling limit
\begin{equation}
q\to 0~~\mbox{and}~~m\to\infty~~~\mbox{with}~~q\,m^{2h_\mathfrak{g}^{\!\vee}}
\equiv \widehat\Lambda^{2h_\mathfrak{g}^{\!\vee}}~~\mbox{fixed}~,
\label{decoupling}
\end{equation}
in which the $\mathcal{N}=2^\star$ theory reduces to the pure $\mathcal{N}=2$ SYM theory. Indeed,
$2h_\mathfrak{g}^{\!\vee}$ is the one-loop $\beta$-function coefficient for the latter.
In this case the one-instanton prepotential is
\begin{equation}
q\,F_{k=1}\Big |_{\mathcal{N}=2}=\widehat\Lambda^{2h_\mathfrak{g}^{\!\vee}}
 \sum_{\alpha\in\Psi^\mathrm{L}_\mathfrak{g}}\frac{1}{(\alpha\cdot a)^2}
\prod_{\beta\in\Psi_\mathfrak{g}(\alpha)}\frac{1}{\beta\cdot a}~.
\label{F1pure}
\end{equation}
This expression perfectly coincides with the known results present in the literature (see for example
\cite{D'Hoker:1996mu} and in particular \cite{Keller:2011ek}), 
while (\ref{F1}) represents the generalization thereof to the $\mathcal{N}=2^\star$ theories with
any gauge algebra $\mathfrak{g}$.

\section{Multi-instanton results from localization}
\label{secn:localization}
For a classical algebra $\mathfrak{g}\in\{{\tilde A}_r,B_r,C_r,D_r\}$ one can efficiently apply 
the equivariant localization methods
\cite{Nekrasov:2002qd,Bruzzo:2002xf,Shadchin:2004yx,Billo:2012st}
to compute the instanton prepotential, order by order in the instanton number $k$. Even if straightforward in principle, these methods become computationally quite involved as $k$ increases, and thus they are practical
only for the first few values of $k$. Nonetheless the information obtained in this way is extremely useful since it provides a benchmark against which one can test the results predicted using the recursion relation and $S$-duality.

The essential ingredient is the instanton partition function
\begin{equation}
Z_k^\mathfrak{g}= 
\oint\prod_{i=1}^{K_{\mathfrak{g}}}\frac{d\chi_i}{2\pi\ii}~z_k^{\mathrm{gauge}}\,z_k^{\mathrm{matter}}
\label{Zkg}
\end{equation}
where $K_{\mathfrak{g}}$ is the number of integration variables given by
\begin{equation}
K_{\mathfrak{g}}=\begin{cases}
~~k &\mbox{for} ~\mathfrak{g}={\tilde A}_r,B_r,D_r~, \\
~\big[\frac{k}{2}\big]&\mbox{for}~\mathfrak{g}=C_r~,
\end{cases}
\label{Kg}
\end{equation}
while $z_k^{\mathrm{gauge}}$ and $z_k^{\mathrm{matter}}$ are, respectively, the contributions 
of the gauge vector multiplet and the matter hypermultiplet in the adjoint representation of $\mathfrak{g}$.
These factors, which are different for the different algebras, depend on the vacuum expectation 
value $a$ and on the deformation parameters $\epsilon_1,\cdots,\epsilon_4$, and are typically 
meromorphic functions of the integration variables
$\chi_i$. The integrals in (\ref{Zkg}) are computed by closing the contours in the upper-half complex
$\chi_i$-planes after giving the $\epsilon$-parameters an imaginary part with the following 
prescription
\begin{equation}
\mathrm{Im}(\epsilon_4)\gg \mathrm{Im}(\epsilon_3)\gg\mathrm{Im}(\epsilon_2)\gg\mathrm{Im}(\epsilon_1)> 0~.
\label{prescription}
\end{equation}
In this way all ambiguities are removed and we obtain the instanton partition function
\begin{equation}
Z_{\mathrm{inst}}^{\mathfrak{g}}=1+\sum_{k\geq1}\,q^k Z_k^{\mathfrak{g}}~.
\end{equation}
At the end of the calculations we have to set
\begin{equation}
\epsilon_3=m-\frac{\epsilon_1+\epsilon_2}{2}~,\quad
\epsilon_4=-m-\frac{\epsilon_1+\epsilon_2}{2}   
\label{mass34}
\end{equation}
in order to express the result in terms of the hypermultiplet mass $m$ in the normalization of the previous sections. Finally, the non-perturbative prepotential of the $\mathcal{N}=2^\star$ SYM theory is given by
\begin{equation}
F_{\mathrm{inst}}^{\mathfrak{g}}=\lim_{\epsilon_1,\epsilon_2\to0}\Bigl(-\epsilon_1\epsilon_2\,\log Z_{\mathrm{inst}}^{\mathfrak{g}}\Bigr)
=\sum_{k\geq1}\,q^k F_k^{\mathfrak{g}}~.
\label{FZ}
\end{equation}
We now provide the explicit expressions of $z_k^{\mathrm{gauge}}$ and $z_k^{\mathrm{matter}}$
for all classical algebras. The details on the derivation of these expressions can be found in \cite{Billo':2015ria,Billo:2013fi} (see also, for example,
\cite{D'Hoker:1996mu} and \cite{Shadchin:2004yx}).

\paragraph*{$\bullet$ The unitary algebras ${\tilde A}_r$~}
In this case the localization techniques yield
\begin{subequations}
\begin{align}
&\!\!\!z_k^{\mathrm{gauge}}=\frac{(-1)^k}{k!} \frac{(\epsilon_1+\epsilon_2)^k }{(\epsilon_1\epsilon_2)^k}
\frac{\Delta(0)\,\Delta(\epsilon_1+\epsilon_2)}{\Delta(\epsilon_1)\,\Delta(\epsilon_2)}\prod_{i=1}^k
\frac{ 1}{P\big(\chi_i\!+\!\frac{\epsilon_1+\epsilon_2}{2}\big)P\big(\chi_i\!-\!\frac{\epsilon_1+\epsilon_2 }{2}\big)}~,\\
\notag\\
&\!\!\!z_k^{\mathrm{matter}}=\frac{ (\epsilon_1+\epsilon_3)^k (\epsilon_1+\epsilon_4)^k}{(\epsilon_3 \epsilon_4)^k}
\frac{\Delta(\epsilon_1+\epsilon_3)\,\Delta(\epsilon_1+\epsilon_4)}{\Delta(\epsilon_3)\,\Delta(\epsilon_4)}
\prod_{i=1}^k P\big(\chi_i\!+\!\ft{\epsilon_3-\epsilon_4}{2}\big)P\big(\chi_i\!-\!\ft{\epsilon_3-\epsilon_4}{2} \big)
\end{align}
\end{subequations}
where
\begin{equation}
P(x) =\prod_{u=1}^{r+1}\big(x- a_u)~,\qquad
\Delta(x)=\prod_{i<j}^k \big(x^2-(\chi_i-\chi_j)^2\big)~.
\end{equation}

\paragraph{$\bullet$ The orthogonal algebras $B_r$ and $D_r$~}
In these cases we find
\begin{subequations}
\begin{align}
&z_k^{\mathrm{gauge}}=\,
\frac{(-1)^k}{2^k\,k!}
\frac{(\epsilon_1+\epsilon_2)^k}{(\epsilon_1\epsilon_2)^k}
\frac{\Delta(0)\,\Delta(\epsilon_1+\epsilon_2)}{\Delta(\epsilon_1)\,\Delta(\epsilon_2)}\,\prod_{i=1}^k
\frac{ 4\chi_i^2 \,\big( 4\chi_i^2-(\epsilon_1+\epsilon_2)^2\big)}{P\big(\chi_i+\frac{\epsilon_1+\epsilon_2}{2}\big)\big(\chi_i-\frac{\epsilon_1+\epsilon_2}{2}\big)}~,\\
\notag\\
&z_k^{\mathrm{matter}}=\frac{ (\epsilon_1+\epsilon_3)^k (\epsilon_1+\epsilon_4)^k}{(\epsilon_3 \epsilon_4)^k}
\frac{\Delta\big(\epsilon_1+\epsilon_3 \big)\Delta\big(\epsilon_1+\epsilon_4 \big)}  {\Delta\big(\epsilon_3 \big)\Delta\big(\epsilon_4 \big)}
\notag\\
&\qquad\qquad\qquad\qquad\qquad\times~
\prod_{i=1}^k
\frac{P\big(\chi_i+\frac{\epsilon_3-\epsilon_4}{2} \big)P\big(\chi_i- \frac{\epsilon_3-\epsilon_4}{2} \big)}{\big( 4 \chi_i^2-\epsilon_3^2 \big)\big( 4 \chi_i^2-\epsilon_4^2 \big)}~,
\end{align}
\end{subequations}
where
\begin{equation}
\begin{aligned}
\Delta(x) &
=\prod_{i<j}^k\big(x^2-(\chi_i-\chi_j)^2)\big)\big(x^2-(\chi_i+\chi_j)^2\big)~,\\
P(x) &= \,x \prod_{u=1}^r\big(x^2-2a_u^2)~\mbox{for}~B_r~,~~~
P(x) = \,\prod_{u=1}^r\big(x^2-a_u^2)~\mbox{for}~D_r~.
\end{aligned}
\end{equation}

\paragraph{$\bullet$ The symplectic algebras $C_r$~}
Finally, for the symplectic algebras we have
\begin{subequations}
\begin{align}
&z_k^{\mathrm{gauge}}=\frac{(-1)^k}{2^{k+\nu} \,k!}\,
\frac{ (\epsilon_1+\epsilon_2)^k}{ (\epsilon_1\epsilon_2)^{k+\nu}}
\,\frac{\Delta(0)\,\Delta(\epsilon_1+\epsilon_2 )}{\Delta(\epsilon_1)\,\Delta(\epsilon_2)}\,
   \frac{1}{P\left(\ft{\epsilon_1+\epsilon_2}{2} \right)^\nu} \\
& \quad\quad\quad \times~\prod_{i=1}^{[\frac{k}{2}]}
\frac{1}{P\big(\chi_i+\frac{\epsilon_1+\epsilon_2 }{2}\big)P\big(\chi_i-\frac{\epsilon_1+\epsilon_2 }{2}\big)  (4\chi_i^2-\epsilon_1^2)   \big( 4\chi_i^2-\epsilon_2^2\big)}~,\notag\\
\notag\\
&z_k^{\mathrm{matter}}=\, 
\frac{(\epsilon_1+\epsilon_3)^{k+\nu} (\epsilon_1+\epsilon_4)^{k+\nu}}{( \epsilon_3 \epsilon_4)^k }
\frac{\Delta\big(\epsilon_1+\epsilon_3 \big)\Delta\big(\epsilon_1+\epsilon_4 \big)  }  {\Delta\big(\epsilon_3 \big)\Delta\big(\epsilon_4 \big)}\,
 P\big(\ft{\epsilon_3-\epsilon_4}{2}\big)^\nu \\
&\quad\quad\quad \times\prod_{i=1}^{[\frac{k}{2}]}
\!P\big(\chi_i+\ft{\epsilon_3-\epsilon_4}{2} \big) P\big(\chi_i- \ft{\epsilon_3-\epsilon_4}{2} \big)
\big( 4 \chi_i^2-(\epsilon_1+\epsilon_3)^2 \big)\big( 4 \chi_i^2-(\epsilon_1+\epsilon_4)^2 \big) ~,
\notag
\end{align}
\end{subequations}
where $\nu=k-2\big[\ft{k}{2}\big]$ and 
\begin{equation}
 \begin{aligned}
P(x) &=  \prod_{u=1}^r\big(x^2-a_u^2)~,  \\
\Delta(x)&=\prod_{i<j}^{[\frac{k}{2}]}\big(x^2-(\chi_i-\chi_j)^2)\big)\big( x^2-(\chi_i+\chi_j)^2\big)
 \prod_{i=1}^{[\frac{k}{2}]} \big( x^2-\chi_i^2\big)^\nu
\end{aligned}
\end{equation}

Using these expressions we have computed the non-perturbative prepotential of the $\mathcal{N}=2^\star$
theories up to $k=5$ for the unitary and simplectic algebras, and up to $k=2$ for the orthogonal algebras.
These explicit results, once rewritten in terms of the root lattice sums (\ref{sumsLS}), are in perfect 
agreement with those obtained using the recursion relation presented in the previous section. This 
agreement provides a highly non-trivial consistency check on the entire construction.

\section{Conclusions}
We have shown that the $S$-duality of $\mathcal{N}=2^\star$ theories 
allows the recursive determination of the terms in the mass expansion of 
the prepotential in terms of (quasi-)modular forms of a suitable subgroup of the $S$-duality group; 
this yields expressions valid at all instanton numbers with very little input from  
microscopic computations. Our results agree with those obtained from localization 
techniques when $\mathfrak{g}$ is a classical algebra 
but, beeing based only on the formal properties of the 
root systems, they represent a solid prediction for the gauge theories based on 
exceptional groups, where no ADHM costruction 
of instantons and no localization methods are avaliable. The original papers
\cite{Billo':2015ria} also discuss the recursion procedure in an $\Omega$-background with generic $\epsilon$ parameters. 

\vskip 1.5cm
\noindent {\large {\bf Acknowledgments}}
\vskip 0.2cm
The work of M.B., M.F. and A.L. is partially supported  by the Compagnia di San Paolo
contract ``MAST: Modern Applications of String Theory'' TO-Call3-2012-0088.

\section*{Appendix}
\label{secn:app}
Here we give our conventions for the root system of all algebras $\mathfrak{g}$ in terms 
of an orthonormal basis $\{\mathbf{e}_i\,;\,1\le i\le \mathfrak{r}\}$ in $\mathbb{R}^{\mathfrak{r}}$
where $\mathfrak{r}=\mathrm{rank}(\mathfrak{g})$.

\paragraph{$\bullet$ ${\tilde A}_r$~}
The roots of ${\tilde A}_r$ are:
\begin{equation}
\big\{\pm (\mathbf{e}_i - \mathbf{e}_j) \,;\,1\le i<j\le r+1\big\}~.
\label{rAn}
\end{equation}
\paragraph{$\bullet$ $B_r$~}
The long and short roots of $B_r$ are, respectively:
\begin{equation}
\left\{\pm \sqrt{2}\,\mathbf{e}_i\pm\sqrt{2}\, \mathbf{e}_j\,;\,1\le i<j\le r\right\}
\quad\mbox{and}\quad\left\{\pm \sqrt{2}\,\mathbf{e}_i\,;\,1\le i\le r\right\}~.
\label{rBn}
\end{equation}
\paragraph{$\bullet$ $C_r$~}
The long and short roots of $C_r$ are, respectively:
\begin{equation}
\left\{\pm2 \,\mathbf{e}_i\,;\,1\le i\le r\right\}
\quad\mbox{and}\quad
\left\{\pm \mathbf{e}_i\pm \mathbf{e}_j\,;\,1\le i<j\le r\right\}~.
\label{rCn}
\end{equation}
\paragraph{$\bullet$ $D_r$~}
The roots of $D_r$ are:
\begin{equation}
\big\{\pm \mathbf{e}_i\pm \mathbf{e}_j\,;\,1\le i<j\le r\big\}~,
\label{rDn}
\end{equation}
\paragraph{$\bullet$ $E_6$~}
The roots of $E_6$ are:
\begin{equation}
\big\{\pm \mathbf{e}_i\pm \mathbf{e}_j\,;\,1\le i<j\le 5\big\}
\cup
\big\{\pm \ft{1}{2}\,\mathbf{e}_1
\cdots \pm \ft{1}{2}\,\mathbf{e}_5
\pm \ft{\sqrt{3}}{2}\, \mathbf{e}_6
\, \big\}~,
\label{rE6}
\end{equation}
where the elements of the second set must have an even number of minus signs. 
\paragraph{$\bullet$ $E_7$~}
The roots of $E_7$ are:
\begin{equation}
\big\{\pm \mathbf{e}_i\pm \mathbf{e}_j\,;\,1\le i<j\le 6\big\}
\cup \big\{\pm\sqrt{2}\,\mathbf{e}_7\big\}
\cup
\big\{\pm \ft{1}{2}\,\mathbf{e}_1
\cdots \pm \ft{1}{2}\,\mathbf{e}_6 \pm \ft{1}{\sqrt{2}}\, \mathbf{e}_7\, \big\}~,
\label{rE7}
\end{equation}
where the elements of the third set must have an odd (even) number of minus signs in the 
$(\mathbf{e}_1, \cdots, \mathbf{e}_6)$ components if the $\mathbf{e}_7$ is positive (negative).
\paragraph{$\bullet$ $E_8$~}
The roots of $E_8$ are:
\begin{equation}
\big\{\pm \mathbf{e}_i\pm \mathbf{e}_j\,;\,1\le i<j\le 8\big\}
\cup
\big\{
\pm \ft{1}{2}\,\mathbf{e}_1\cdots  \pm \ft{1}{2}\,\mathbf{e}_8
\, \big\}~,
\label{rE8}
\end{equation}
where the element of the second set must have an even number of minus signs.
\paragraph{$\bullet$ $F_4$~}
The long roots of $F_4$ are:
\begin{equation}
\left\{\pm \sqrt{2}\,\mathbf{e}_i\pm\sqrt{2}\, \mathbf{e}_j\,;\,1\le i<j\le 4\right\}~,
\label{longF4}
\end{equation}
while the short roots are:
\begin{equation}
\left\{\pm \sqrt{2}\,\mathbf{e}_1,\pm \sqrt{2}\,\mathbf{e}_2,\pm \sqrt{2}\,\mathbf{e}_3,\pm \sqrt{2}\,\mathbf{e}_4,\,\pm\ft{1}{\sqrt{2}}\,\mathbf{e}_1\pm\ft{1}{\sqrt{2}}\,\mathbf{e}_2
\pm\ft{1}{\sqrt{2}}\,\mathbf{e}_3 \pm\ft{1}{\sqrt{2}}\,\mathbf{e}_4\right\}~.
\label{shortF4}
\end{equation}
\paragraph{$\bullet$ $G_2$~}
The long and short roots of $G_2$ are, respectively:
\begin{equation}
\left\{\pm \ft{3}{\sqrt{2}}\,\mathbf{e}_1\pm\sqrt{\ft{3}{2}}\,\mathbf{e}_2\,,\,\pm\sqrt{6}\,\mathbf{e}_2\right\}
\quad\mbox{and}\quad
\left\{\pm\sqrt{2}\,\mathbf{e}_1\,,\,
\pm \ft{1}{\sqrt{2}}\,\mathbf{e}_1\pm\sqrt{\ft{3}{2}}\,\mathbf{e}_2\right\}~.
\label{rG2}
\end{equation}

\vspace{10pt}
Finally, in the following table we collect the main properties for the various algebras that are useful for the calculations presented in the main text:

\vspace{5pt}
\hspace{-10pt}
\begin{tabular}{c|c|c|c|c|c|c|c}
{\small{$\mathfrak{g}\phantom{\Big|}$}}
&{\small{$\mathrm{dim}$}}
&{\small{$\mathrm{rank}$}}
&$h^{\!\vee}$
&{\small{$\mathrm{ord}\big(\Psi^{\mathrm{L}}_{\mathfrak{g}}\big)$}}
&{\small{$\mathrm{ord}\big(\Psi^{\mathrm{S}}_{\mathfrak{g}}\big)$}}
&{\small{$\mathrm{ord}\big( \Psi_{\mathfrak{g}}(\alpha_{\mathrm{L}})\big)$}}
&{\small{$\mathrm{ord}\big( \Psi_{\mathfrak{g}}^{\!\vee}(\alpha_{\mathrm{S}})\big)$}}
\\
\hline
\hline
{\small{$A_{r}\phantom{\Big|}\!\!$}}
&{\small{$(r+1)^2$}}
&{\small{$r+1$}}
&{\small{$r+1$}}
&{\small{$r(r+1)$}}
&{\small{--}}
&{\small{$2r-2$}}
&{\small{--}}\\
\hline
{\small{$B_r\phantom{\Big|}\!\!$}}
&{\small{$r(2r+1)$}}
&{\small{$r$}}
&{\small{$2r-1$}}
&{\small{$2r(r-1)$}}
&{\small{$2r$}}
&{\small{$4r-6$}}
&{\small{$2r-2$}}\\
\hline
{\small{$C_r\phantom{\Big|}\!\!$}}
&{\small{$r(2r+1)$}}
&{\small{$r$}}
&{\small{$r+1$}}
&{\small{$2r$}}
&{\small{$2r(r-1)$}}
&{\small{$2r-2$}}
&{\small{$4r-6$}}\\
\hline
{\small{$D_r\phantom{\Big|}\!\!$}}
&{\small{$r(2r-1)$}}
&{\small{$r$}}
&{\small{$2r-2$}}
&{\small{$2r(r-1)$}}
&{\small{--}}
&{\small{$4r-8$}}
&{\small{--}}\\
\hline
{\small{$E_6\phantom{\Big|}\!\!$}}
&{\small{$78$}}
&{\small{$6$}}
&{\small{$12$}}
&{\small{$72$}}
&{\small{--}}
&{\small{$20$}}
&{\small{--}}\\
\hline
{\small{$E_7\phantom{\Big|}\!\!$}}
&{\small{$133$}}
&{\small{$7$}}
&{\small{$18$}}
&{\small{$126$}}
&{\small{--}}
&{\small{$32$}}
&{\small{--}}\\
\hline
{\small{$E_8\phantom{\Big|}\!\!$}}
&{\small{$248$}}
&{\small{$8$}}
&{\small{$30$}}
&{\small{$240$}}
&{\small{--}}
&{\small{$56$}}
&{\small{--}}\\
\hline
{\small{$F_4\phantom{\Big|}\!\!$}}
&{\small{$52$}}
&{\small{$4$}}
&{\small{$9$}}
&{\small{$24$}}
&{\small{$24$}}
&{\small{$14$}}
&{\small{$14$}}\\
\hline
{\small{$G_2\phantom{\Big|}\!\!$}}
&{\small{$14$}}
&{\small{$2$}}
&{\small{$4$}}
&{\small{$6$}}
&{\small{$6$}}
&{\small{$4$}}
&{\small{$4$}}\\
\end{tabular}
 

\end{document}